\documentclass[12pt]{iopart}

\begin{document}

\title{Small Cosmological Constants from a Modified Randall-Sundrum Model}

\author{Bei Jia$^{1,2}$, Xi-Guo Lee$^{1,3}$}

\address{$^1$Institute of Modern Physics,
Chinese Academy of Sciences, P.O.Box 31 Lanzhou, 730000,
China

$^2$Graduate University of Chinese Academy of Sciences, Beijing,
100080, China

$^3$Center of Theoretical Nuclear Physics, National Laboratory of
Heavy Ion Collisions, P.O.Box 31 Lanzhou, 730000, China}

\begin{abstract}
We study a mechanism, inspired from the mechanism for generating the
gauge hierarchy in Randall-Sundrum model, to investigate the
cosmological constant problem. First we analyze the bulk
cosmological constant and brane vacuum energies in RS model. We show
that the five-dimensional bulk cosmological constant and the vacuum
energies of the two branes all obtain their natural values. Finally
we argue how we can generate a small four-dimensional effective
cosmological constant on the branes through modifying the original
RS model.

\end{abstract}

\maketitle

\section{Introduction}
Since the early works of Kluza and Klein [1] about unifying
gravitational and electromagnetic interactions based on
five-dimensional spacetime, the topic of extra dimensions and their
physical implications has been considered by numerous physicists.
The main topic of models with extra dimensions in recent years has
been in the form of ``brane world'' scenarios, in which all matter
fields are confined to a ``brane'' while only gravity propagates
along the extra dimensions, with different purposes of solving
problems ranging from hierarchy problem to symmetry breaking.
Randall-Sundrum (RS) model [2][3] aims at generating the hierarchy
between the TeV scale and the Planck scale, with a five-dimensional
mtric in an $AdS_5$ spacetime

\begin{equation} \label{eps}
ds^2=e^{-2kr_{c}|\phi|}\eta_{\mu\nu}dx^{\mu}dx^{\nu}+r^{2}_{c}d\phi^{2}
\end{equation}
The extra dimension is compactified as $S^1/\mathbf{Z}_2$, and there
are two 3-branes which locate at the orbifold fixed points
$\phi=0,\pi$. The term $e^{-2kr_{c}|\phi|}$ is called ``warped
factor''. The action in RS model is

\begin{equation} \label{eps}
S=\int d^4x dy \sqrt{-g}\ [2M^3R-\Lambda]+\sum_{i=1,2}\int d^4x
\sqrt{-g^{(i)}}[\mathcal {L}_i-\Lambda_i]
\end{equation}
This action includes a five-dimensional cosmological constant
$\Lambda$ and two brane vacuum energies denoted by $\Lambda_i$.
$\mathcal {L}_i$ represent the lagrangian of matter fields on the
branes, while $g^{(i)}_{\mu\nu}$ represent the induced metric on the
branes. Randall and Sundrum showed that the four-dimensional Planck
mass $M_{Pl}$ can be related to the five-dimensional fundamental
mass scale $M$ as

\begin{equation} \label{eps}
M_{Pl}^2=\frac{M^3}{k}[1-e^{-2kr_c\pi}]
\end{equation}
where $r_c$ represents the length of the fifth dimension. This means
$M_{Pl}$ depends only weakly on $r_c$ in the large $kr_c$ limit,
which is completely different other models involving extra
dimensions. Therefore, Randall and Sundrum assumed that
$M_{Pl}\approx M$ as the fundamental mass scale. It is also shown in
RS model that a hierarchy between a physical mass parameter $m$ and
a fundamental mass parameter $m_{0}$ can be naturally generated

\begin{equation} \label{eps}
m=e^{-kr_{c}\pi}m_{0}
\end{equation}
If $e^{kr_{c}\pi}$ is of order $10^{15}$, then this mechanism
produces TeV physical mass scales from fundamental mass parameters
not far from the Planck scale. In the seconde paper [3] RS discussed
how to recover four-dimensional gravity through using an infinite
uncompactified extra dimension.

The observed small value of the cosmological constant raised the so
called cosmological constant problem [4]. A straightforward analyze
shows that this constant should be of order $M_{Pl}^{4}$, which is
nearly 120 times of magnitude larger than its observed value [5].
There are many works on solving the cosmological constant problem,
such as works based on supergravity [6] and string theory [7]. In
the brane-world models, if there is four-dimensional sources in
five-dimensional spacetime, the effects of the four-dimensional
brane sources can be balanced by a five-dimensional cosmological
constant to get a theory in which the effective cosmological
constant on the brane would be vanishing. This leads to the so
called ``warped'' extra dimensions. There are many attempts of
solving the cosmological constant problem through brane-world models
[8].

In this paper we first analyze the vacuum energies and cosmological
constant in the conventional RS model. We show that the
five-dimensional cosmological constant and the four-dimensional
cosmological constants of the two branes are of their ``natural''
value: $-\Lambda\sim M^{5}_{Pl}$ and $-V_{vis}(=V_{hid})\sim
M^{4}_{Pl}$. Then in section 3 we discuss the cosmological constant
problem, without concerning the hierarchy problem .We analyze a
mechanism which can generate a small four-dimensional effective
cosmological constant on the brane, not the gauge hierarchy, through
modifying the original RS model.

\section{Bulk and Brane Cosmological Constants in RS model}
It is straightforward to analyze the vacuum energies in RS model. In
RS model the four-dimensional Planck mass can be expressed as [2]

\begin{equation} \label{eps}
M^2_{Pl}=\frac{M^3}{k}[1-e^{-2kr_c\pi}],\qquad
k=\sqrt{\frac{-\Lambda}{24M^3}}
\end{equation}

In order to solve the hierarchy problem, according to RS,
$e^{kr_c\pi}$ should be of order $10^{15}$. Then

\begin{equation} \label{eps}
M^2_{Pl}\sim\frac{M^3}{k}= M^3\sqrt{\frac{24M^3}{-\Lambda}}
\end{equation}
In RS model, the fundamental mass scale is set by identifying the
four-dimensional Planck mass $M_{Pl}$ with the five-dimensional
scale $M$, then we have

\begin{equation} \label{eps}
-\Lambda\sim M^5_{Pl}
\end{equation}
which is the natural value of $\Lambda$ as the five-dimensional
cosmological constant. On the other hand, in RS model the vacuum
energies, or brane tensions, of the two branes, are determined by
the five-dimensional cosmological constant $\Lambda$ and the
five-dimensional mass scale $M$

\begin{equation} \label{eps}
-V_{vis}=V_{hid}=24kM^3
\end{equation}
Again, if we set $M_{Pl}\sim M$, we can get

\begin{equation} \label{eps}
-V_{vis}=V_{hid}\sim M^4_{Pl}
\end{equation}
which is also natural because $V_{vis}$ and  $V_{hid}$ are the
four-dimensional vacuum energies on the visible and hidden 3-branes.
Therefore, in RS model the five-dimensional and four-dimensional
vacuum energies reach their natural values if the mechanism of
solving the hierarchy problem works well.

In RS model the only required fine-tuning is $kr_c\sim10$, so that
the magnitude of the extra dimension is

\begin{equation} \label{eps}
r_c\sim 10^{18} \textrm{GeV}^{-1} \sim 10^{-34} \textrm{m}
\end{equation}
which is of the magnitude of the Planck length.

\section{A Small Cosmological Constant}

Notice that in RS model the vacuum energies of the two branes
already exist in the setup of the action (2). Here we consider the
condition that there are no pre-existing vacuum energies on the
branes; the four-dimensional effective cosmological constants on the
branes are induced by the five-dimensional bulk cosmological
constant. This change does not affect the form of the warped factor
in RS model, so Equations (5) still hold (notice that the brane
vacuum energies do not exit here, so the discussion about them do
not hold anymore). However, we will need an extremely small $r_c$
here, as well as the constrain that $M_{Pl}\sim M$, so that from
Equation (5) we have

\begin{equation} \label{eps}
-\Lambda\sim M_{Pl}^5(1-e^{-2kr_c\pi})^2=M_{Pl}^5\varepsilon^2
\end{equation}
which is different from Equation (7). Here we require that
$\varepsilon$ is extremely small. The reason for this requirement
will be seen latter.

Like in RS model, the fifth dimension is $S^1/\mathbf{Z}_2$. We
start with the action of 5D gravity

\begin{equation} \label{eps}
S=\int d^4x\ \int ^{\pi r_c}_{-\pi r_c}dy\sqrt{-G}(-\Lambda+2M^3R)
\end{equation}
and the metric

\begin{equation} \label{eps}
ds^2=e^{-2k|y|}\eta_{\mu\nu}dx^{\mu}dx^{\nu}+dy^{2}
\end{equation}
where $y=r_c\phi$ represents the extra dimension.

In order to induce the four-dimensional effective cosmological
constant, we integrate over the fifth dimension $y$

\begin{equation} \label{eps}
S_\Lambda=\int d^4x\ \sqrt{-6M^3\Lambda}(1-e^{-4kr_c\pi})
\end{equation}
If we refer the four-dimensional mass scale $M_{Pl}$ as the
fundamental mass scale, i. e. if we set $M_{Pl}\sim M$, then
according to the our previous discussion, the bulk cosmological
constant take the value in Equation (11), so that

\begin{equation} \label{eps}
S_\Lambda\sim\int d^4x\ M_{Pl}^4\varepsilon(1-e^{-4kr_c\pi})
\end{equation}
Compare this result to the four-dimensional effective action on the
brane at $y=r_c\pi$, whose induced metric is
$g^{vis}_{\mu\nu}=e^{-2kr_c\pi}\eta_{\mu\nu}$

\begin{equation} \label{eps}
S_{eff}=\int d^4x\
\sqrt{-g_{vis}}(-\Lambda^{vis}_{(4)}+2M^2_{Pl}R^{vis}_{(4)})
\end{equation}
we get

\begin{equation} \label{eps}
-\Lambda^{vis}_{(4)}\sim
M^4_{Pl}(1-e^{-2kr_c\pi})(e^{4kr_c\pi}-1)\sim
M^4_{Pl}\varepsilon\delta
\end{equation}
We can see, if $\varepsilon\delta$ is of order $10^{-120}$, then we
can generate a small 4D effective cosmological constant,
$-\Lambda^{vis}_{(4)}\sim 10^{-47}\textrm{GeV}^{4}$, with a slightly
larger five-dimensional cosmological constant (11) in the bulk.

Recall that in RS model, in order to generate a exponential
hierarchy between TeV scale and Planck scale, they require $kr_c\sim
10$ so that $r_c\sim 10^{-34} \textrm{m}$. Now in order to generate
the huge ``hierarchy'' between the natural value and observed value
of the cosmological constant, $r_c$ has to be extremely small. In
this case, the bulk cosmological constant and the vacuum energies on
the two branes are all very small due to the smallness of $r_c$.

Notice that this small $r_c$ would be too small to solve the gauge
hierarchy problem. However, since our concern here is the
cosmological constant problem, not the hierarchy problem, this
result only means that we cannot solve both of the two problems at
the same time in this modified model. Also, this induced 4D
effective cosmological constant is on the ``visible'' brane which
locates at $y=r_c\pi$. Similarly, for the ``hidden'' brane at $y=0$,
the effective action is (the induced metric on this brane is
$g^{hid}_{\mu\nu}=\eta_{\mu\nu}$)

\begin{equation} \label{eps}
S_{eff}=\int d^4x\
\sqrt{-g_{hid}}(-\Lambda^{hid}_{(4)}+2M^2_{Pl}R^{hid}_{(4)})
\end{equation}
we find that on this brane the induced cosmological constant can be
again very small

\begin{equation} \label{eps}
-\Lambda^{hid}_{(4)}\sim M^4_{Pl}\varepsilon(1-e^{-4kr_c\pi})
\end{equation}
This means that if $r_c$ is extremely small, we can naturally obtain
a small effective cosmological constant on both of the two branes.

\section{Conclusion}
The mechanism of solving the gauge hierarchy problem in RS model can
be used to investigate the cosmological constant problem, if we make
some changes to the original RS model. We show that in the original
RS model the five-dimensional bulk cosmological constant and the
vacuum energies of the two branes obtain their natural value.
Through assuming that the four-dimensional cosmological constants on
the two branes are the effective vacuum energies induced by the bulk
cosmological constant, we are able to obtain two naturally small 4D
effective cosmological constants on the two branes from a small
five-dimensional cosmological constant in the bulk. This mechanism
will put a more restrictive constrain on the size of the extra
dimension than the original RS model, that $r_c$ has to be extremely
small. However, like in RS model, this is the only requirement in
order to generate a small cosmological constant.

\ack{This work is supported by the CAS Knowledge Innovation Project
(No.KJCX3-SYW- N2,No.KJCX2-SW-N16) and the Science Foundation of
China (10435080, 10575123).}

\maketitle

\end{document}